\documentclass[aps,prc,twocolumn,groupedaddress,showpacs]{revtex4}
\usepackage{graphicx}

\begin{document}
\newcommand{\opg}{\mbox{$^{16}$O(p,$\gamma)^{17}$F}}
\newcommand {\nc} {\newcommand}
\nc {\beq} {\begin{eqnarray}}
\nc {\eol} {\nonumber \\}
\nc {\eeq} {\end{eqnarray}}
\nc {\ve} [1] {\mbox{\boldmath $#1$}}
\nc {\vS} {\mbox{$\ve{S}$}}


\title{New reaction rate for $^{16}$O(p,$\gamma$)$^{17}$F and its influence on the oxygen isotopic ratios in massive AGB stars}


\author{C. Iliadis$^{1,2}$, C. Angulo$^{3}$, P. Descouvemont$^{4}$, M. Lugaro$^{5,6}$ and P. Mohr$^{7}$}
\affiliation{$^{1}$Department of Physics and Astronomy, University of North Carolina, Chapel Hill, North Carolina, 27599-3255, USA}
\affiliation{$^{2}$Triangle Universities Nuclear Laboratory, Durham, North Carolina 27708-0308, USA}
\affiliation{$^{3}$Tractebel Engineering (SUEZ), Avenue Ariane 7, 1200 Brussels, Belgium}
\affiliation{$^{4}$Physique Nucl\'{e}aire Th\'{e}orique et Physique Math\'{e}matique, CP229, Universit\'{e} Libre de Bruxelles, B-1050 Brussels, Belgium}
\affiliation{$^{5}$Sterrenkundig Instituut, University of Utrecht, Postbus 80000, 3508 TA Utrecht, The Netherlands}
\affiliation{$^{6}$Centre for Stellar and Planetary Astrophysics, School of Mathematical Sciences, Monash University, Victoria 3800, Australia}
\affiliation{$^{7}$Diakoniekrankenhaus Schw\"{a}bisch Hall, D-74523 Schw\"{a}bisch Hall, Germany}

\date{\today}

\begin{abstract}
{
The $^{16}$O(p,$\gamma$)$^{17}$F reaction rate is revisited with
special emphasis on the stellar temperature range of T=60-100 MK
important for hot bottom burning in asymptotic giant branch (AGB)
stars. We evaluate existing cross section data that were
obtained since 1958 and, if appropriate, correct published data for
systematic errors that were not noticed previously, including the
effects of coincidence summing and updated effective stopping
powers. The data are interpreted by using two different models
of nuclear reactions, that is, a potential model and
R-matrix theory. A new astrophysical S-factor and recommended
thermonuclear reaction rates are presented. As a result of our work,
the $^{16}$O(p,$\gamma$)$^{17}$F reaction has now the most precisely known rate involving any target nucleus in the mass $A\geq12$ range, with reaction rate errors
of about 7\% over the entire temperature region of astrophysical interest (T=0.01-2.5 GK). The impact of the present improved reaction rate with its significantly reduced uncertainties on the hot bottom burning in AGB stars is discussed. In contrast to earlier results we find now that there is not clear evidence to date for any stellar grain origin from massive AGB stars.
}
\end{abstract}

\maketitle

\section{Introduction}

The $^{16}$O(p,$\gamma$)$^{17}$F reaction is characterized by a number
of exceptional attributes. At lower bombarding energies it provides a
textbook example for a nonresonant reaction cross section since the
lowest-lying resonance is located at a relatively high laboratory
energy of 2.66 MeV \cite{TIL93}. The absence of low-energy resonances
and the high binding energy of $^{16}$O are among the main reasons for
the fact that the cross section can be described in terms of simple
nuclear reaction models. Indeed, the $^{16}$O(p,$\gamma$)$^{17}$F
reaction is a prime example for the direct capture reaction model
which assumes that the projectile is captured via a single-step
process into a final state orbit outside a closed $^{16}$O core (see,
for example, Ref. \cite{ROL73}). The absence of low-energy resonances
is also the reason for the fact that the $^{16}$O(p,$\gamma$)$^{17}$F
reaction is the slowest process among all the proton-induced reactions
in the CNO target mass region \cite{ILI07}. This reaction has been
studied many times at low energies. The experimental techniques
applied include the activation method \cite{HES58,TAN59}, the in-beam
study of prompt $\gamma$ -rays \cite{ROL73,CHO75,MOR97} and
measurements in inverse kinematics \cite{BEC82}. It is generally
assumed that the different measurements are in agreement. Despite 
these facts, the thermonuclear reaction rates
evaluated by the NACRE collaboration have relatively large errors
amounting, for example, to $+$35\% and $-$43\% at T=0.06-0.1 GK \cite{ANG99}.

Many stars, including the Sun, will eventually pass through an
evolutionary phase which is referred to as the asymptotic giant
branch (AGB; see Ref. \cite{Her05} for a review). This phase involves a hydrogen and a helium shell that burn
alternately surrounding an inactive stellar core. Convection
carries the products of nucleosynthesis to the stellar
surface where material is ejected via strong stellar winds
\cite{HO04}. A fraction of the ejected matter condenses in form of small grains. Some
of the grains travelled the interstellar medium and have been incorporated into primitive meteorites at the birth of the solar system. Such stellar grains can be extracted in the laboratory from their host meteorites and their isotopic composition can be analyzed
with high precision \cite{LUG05}. The measured isotopic abundance ratios deviate
substantially from solar system abundance ratios, which result from a
homogeneous mixture of contributions from countless stars prior to solar
system formation. Stellar grains, instead, exhibit isotopic abundance
ratios that are characteristic for the composition of their parent
stars, and hence they most likely provide strong constraints for stellar
models. The present work has been motivated by studies showing that variations of the
$^{16}$O(p,$\gamma$)$^{17}$F reaction rate influence sensitively the
$^{17}$O/$^{16}$O isotopic ratio predicted by models of
massive ($\geq4M_{\odot}$) AGB stars, where proton captures occur at the base of the convective envelope (hot bottom burning, HBB). Specifically, a
recent study demonstrated that a fine-tuning of the
$^{16}$O(p,$\gamma$)$^{17}$F reaction rate may account for the
measured anomalous $^{17}$O/$^{16}$O abundance ratio in the
extraordinary presolar spinel grain OC2 \cite{LUG07}.

In the present work we focus our attention on a detailed evaluation of
existing $^{16}$O(p,$\gamma$)$^{17}$F cross section data
(Sec. II). This level of detail was not practical in the broad NACRE
compilation which includes many different reactions. We interpret the
evaluated data in terms of two different nuclear reaction models (Sec. III)
and extract astrophysical S-factors, reaction rates and
associated errors (Sec. IV). The new $^{16}$O(p,$\gamma$)$^{17}$F
thermonuclear reaction rates are then incorporated into stellar models
in order to derive improved oxygen isotopic ratios from the hot bottom
burning in AGB stars (Sec. V).

\section{Evaluation of existing data}\label{dta}
\begin{figure}[]
\includegraphics[height=6cm]{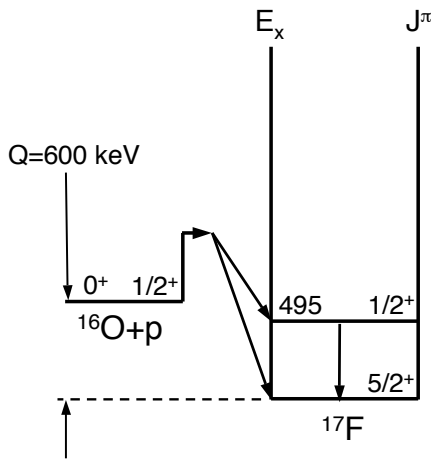}
\caption{\label{fig:leveldiagram} 
Nuclear energy level diagram for
$^{16}$O(p,$\gamma$)$^{17}$F. The level parameters and the reaction
Q-value are adopted from Ref. \cite{TIL93} and Ref. \cite{AUD03},
respectively.  
}
\end{figure}

The radiative proton capture on $^{16}$O (Q=600.27$\pm$0.25 keV
\cite{AUD03}) can proceed via E1 radiation either to the ground state (J$^\pi$=5/2$^+$)
or to the first excited state (E$_x$=495.33$\pm$0.10 keV; J$^\pi$=1/2$^+$) of
$^{17}$F \cite{TIL93}. A nuclear level diagram is shown in
Fig. \ref{fig:leveldiagram}. In this section, we evaluate the existing
cross section data and discuss certain corrections that we deem
necessary. Notice that we make a distinction between experiments that
measure separately the transitions to the two final states and those
that report only on the sum contribution of both transitions. The data
will be displayed in terms of the astrophysical S-factor, defined as
\begin{equation}
{
S(E) = E~\sigma(E)~e^{2\pi\eta}}
\label{eqn:sfactor}
\end{equation}
with $\sigma$ the reaction cross section, $e^{-2\pi\eta}$ the Gamow
factor, and $\eta$ the Sommerfeld parameter. Throughout this experimental section, all
kinematic quantities are given in the laboratory system, unless
mentioned otherwise.

\subsection{Data of Chow, Griffiths and Hall (1975)}

The work of Chow, Griffiths and Hall \cite{CHO75} reports on cross
sections for $^{16}$O(p,$\gamma$)$^{17}$F at seven bombarding energies
in the range of E$_p$=845--2554 keV. The cross sections are separately
given in their Tab. 4 for the transitions to the ground state (4 data points) and first
excited state (7 data points) of $^{17}$F. It is important to emphasize that these
authors normalized the radiative capture cross section to the
Rutherford scattering cross section at E$_p$=405 keV. This procedure
provides capture cross sections independently from the target
stoichiometry, stopping powers and beam straggling and thus avoids
many normalization errors that are frequently associated with absolute
cross section determinations. For more information, see
Ref. \cite{ILI07}.

Not all the experimental details are provided in Ref. \cite{CHO75} in
order to assess the quoted errors. Some of their reported cross
section errors are as small as $\pm$3\%. However, the elastic
scattering measurement alone contributes an error of about $\pm$2\%,
as can be seen from their Tab. 1. Based on the geometry of the
experiment and the methods applied, the following estimates for
additional error contributions seem reasonable: $\pm$3\% for the
$\gamma$-ray efficiency; $\pm$1\% for the escape peak detection;
$\pm$1\% for the angular distribution correction; and $\pm$3\% for the
determination of the effective bombarding energy in their 25 keV thick
solid oxygen targets. Other sources of errors, such as corrections for
the finite solid angle of the $\gamma$-ray detector, are mentioned in
Ref. \cite{CHO75}. The quadratic addition of the above mentioned
errors gives a value of $\pm$5\% and, therefore, we suspect that the
smallest of the capture cross section errors quoted in their Tab. 4
have been somewhat underestimated. As a result of the above
discussion, we adopt errors of at least $\pm$5\% for their quoted
capture cross sections. The modified data of Chow, Griffiths and Hall
\cite{CHO75} are displayed in Fig. 2.

\subsection{Data of Morlock et al. (1997) \label{mormor}}
The measurement of Morlock et al. \cite{MOR97} (see also
Ref. \cite{MOR96}) was performed in the energy range of
E$_p$=280--3740 keV by using an extended gas target. Their
$^{16}$O(p,$\gamma$)$^{17}$F cross section was determined relative to
the $^{16}$O(p,p)$^{16}$O elastic scattering cross section which, in
turn, was normalized to the Rutherford scattering of protons on heavy
noble gases (krypton and xenon). A number of effects were considered
by Refs. \cite{MOR97,MOR96} in the analysis, including the energy loss
and straggling of protons in the gas target, the energy spread of the
beam, the finite solid angle of the $\gamma$-ray detector, the angular
distribution of the emitted $\gamma$-rays, and the dependence of the
detection efficiency on $\gamma$-ray energy and emission angle.

We noticed, however, that the data of Refs. \cite{MOR97,MOR96} were
not corrected for coincidence summing. This effect arises since the
primary $\gamma$-ray populating the E$_x$=495 keV level and the
secondary $\gamma$-ray from the subsequent de-excitation of this state
are in time-coincidence (Fig. 1). Since both photons are emitted
nearly simultaneously, the detector is not able to distinguish between
the two, but instead records a single event with the summed
energy. This experimental artifact gives rise to too many counts for
the ground state transition (``summing-in"), and too few counts for
the transition to the first excited $^{17}$F state
(``summing-out"). The effect can only be corrected for with the proper
knowledge of both the peak and total detection efficiencies (see
Ref. \cite{ILI07} for details). Most of these quantities have not been
reported in Refs. \cite{MOR97,MOR96}, but can be deduced from other
sources, as described below.

We adopted the peak efficiencies from Ref. \cite{MOR96} and from
Refs. \cite{KOE99,KOE98}. The latter studies were performed with a
similar detection setup as in Morlock et al. \cite{MOR97}. In order to
estimate the total detection efficiency, we performed GEANT
simulations for a similar setup involving a Germanium $\gamma$-ray
detector of 100\% relative efficiency. We find summing-out
correction factors of 1.06$\pm$0.02 for the primary transition to the
first excited $^{17}$F state and 1.05$\pm$0.02 for the secondary
de-excitation of the 495 keV level. For the primary ground state
transition, the summing-in depends on the cross section of the transition DC$\rightarrow$495. The correction reduces the reported cross
section at the lowest measured energies (few 100 keV) by about 10\%,
while the correction is negligible at the highest energies (several
MeV).

We found evidence in Morlock's thesis \cite{MOR96} that an additional error of 10\% has to be applied to all data points. Note that this additional error was disregarded in Ref. \cite{MOR97}. We also discovered an inaccuracy (too few significant digits) in the numerical conversion from cross section to S--factor presented in Ref. \cite{MOR96}. The modified data of Morlock et al. \cite{MOR97} and Morlock \cite{MOR96} are displayed in Fig. 2.

\subsection{Data of Hester, Pixley and Lamb (1958)}
Hester, Pixley and Lamb \cite{HES58} measured the total
$^{16}$O(p,$\gamma$)$^{17}$F cross section at six energies in the
bombarding energy range of $E_p$=140--170 keV. They measured the
$^{17}$F activity after exposing an infinitely thick oxygen target to
an intense proton beam. Thus they could not distinguish between
transitions to the ground or first excited state in $^{17}$F. Their
study represents an absolute cross section measurement, in the sense
that the deduced cross section depends on absolute detection
efficiencies, absolute incident charge integration, stopping powers
and target stoichiometry. Recall that many of the systematic errors
related to these sources are avoided in the studies of
Refs. \cite{CHO75,MOR97} mentioned above.

The cross section of Ref. \cite{HES58} was obtained from the expression
\begin{equation}
{
\sigma(E)=\frac{Z_t \epsilon_{eff} Y}{2E^{3/2}}\left( 1+\frac{\sqrt{E}}{Z_t} + ... \right)
}
\label{eqn:hesterexpr}
\end{equation}
with $Z_t$ the charge of the target nucleus, $E$ the laboratory
bombarding energy in MeV, $Y$ the thick target yield and
$\epsilon_{eff}$ the effective stopping power. This expression is
derived under the following assumptions: (i) the target is infinitely
thick; (ii) the S-factor is approximately constant in the energy range
of the measurement; (iii) the stopping power is constant over the
effective thickness of the target. For the effective stopping power of
their Al$_2$O$_3$ target,
$\epsilon_{eff}=\epsilon_O+(2/3)\epsilon_{Al}$, they used a value of
$\epsilon_{eff}=2.8\times10^{-14}$ eVcm$^2$ that was adopted from an
unpublished report \cite{WHA}. We corrected the measured cross
sections of Ref. \cite{HES58} by using modern stopping powers derived
from the code SRIM \cite{SRIM}. The correction decreases the cross
section reported in Ref. \cite{HES58} by 8\% (2\%) at the highest
(lowest) energy measured. The modified data of Hester, Pixley and Lamb
\cite{HES58} are displayed in Fig. 2.

\subsection{Data of Becker et al. (1982)}
Becker et al. \cite{BEC82} measured the cross section for radiative
capture on $^{16}$O at a center-of-mass bombarding energy of
E$_p^{cm}$=853 keV. Their study was performed in inverse kinematics by
directing a $^{16}$O heavy-ion beam onto an extended windowless
hydrogen gas target. It must be emphasized that Becker et
al. performed similar measurements for other nuclear reactions (see
Tab. 1 in Ref. \cite{BEC82}) and that their reported resonance
strengths are frequently used in the literature as absolute strength
standards (see, for example, Refs. \cite{ILI07,CHA07,NEW07}). For the
$^{16}$O(p,$\gamma$)$^{17}$F cross section at E$_p^{cm}$=853 keV they
report a value of $\sigma=(0.92\pm0.13)\times10^{-3}$ mb. Although not
clearly stated in their paper, this value does not correspond to the
total cross section but represents the cross section for the
transition to the first excited $^{17}$F state only \cite{BEC07}. As
will be shown later, the data point of Becker et al. \cite{BEC82}
agrees with the data of Chow, Griffiths and Hall \cite{CHO75} and
Morlock et al. \cite{MOR97}.

\subsection{Other data}

Two more studies \cite{TAN59,ROL73} report low-energy cross sections
for the $^{16}$O(p,$\gamma$)$^{17}$F reaction. These report total
cross sections and provide no information on individual transitions to
the two final $^{17}$F states. We disregarded these data for the
reasons given below.

The study by Tanner \cite{TAN59} reports cross sections in the
bombarding energy range of E$_p$=274--616 keV. The absolute cross
section measurement was performed with a WO$_3$ target by using the
activation method. At the time, stopping powers for tungsten were not
available and, as the closest approximation, tabulated values for
tantalum were used instead \cite{BAD56}. Clearly, these stopping
powers are several decades old and the cross sections reported by
Tanner \cite{TAN59} should in principle be corrected by using modern
stopping power values for oxygen and tungsten. However, our attempt at
a correction was futile, mainly because Tanner obtained the cross
section from the yield and the effective stopping powers of his 150 keV thick target  
by numerical integration. Since the measured yields are not provided
in Ref. \cite{TAN59} there is no obvious way to correct his reported cross
sections. Consequently, we are compelled to disregard the data of Tanner \cite{TAN59} in our analysis.

Rolfs \cite{ROL73} reports total $^{16}$O(p,$\gamma$)$^{17}$F cross
sections in the bombarding energy range of E$_p$=0.3--3.0 MeV. He did
not determine the absolute scale of the cross section, but normalized
his results relative to the absolute cross section of Tanner
\cite{TAN59} that was measured at an energy of E$_p$=616 keV. Since we
did not succeed in correcting the latter data for improved stopping
powers (see above), we are also compelled to disregard the data of
Rolfs \cite{ROL73} in our analysis. 
Nevertheless, the data of Ref. \cite{ROL73}, which were measured over a broad
energy range, demonstrated the rise of the S--factor at low
energies and the measured energy dependence agrees with later
results \cite{MOR97}.

\begin{figure}[]
\includegraphics[height=10cm]{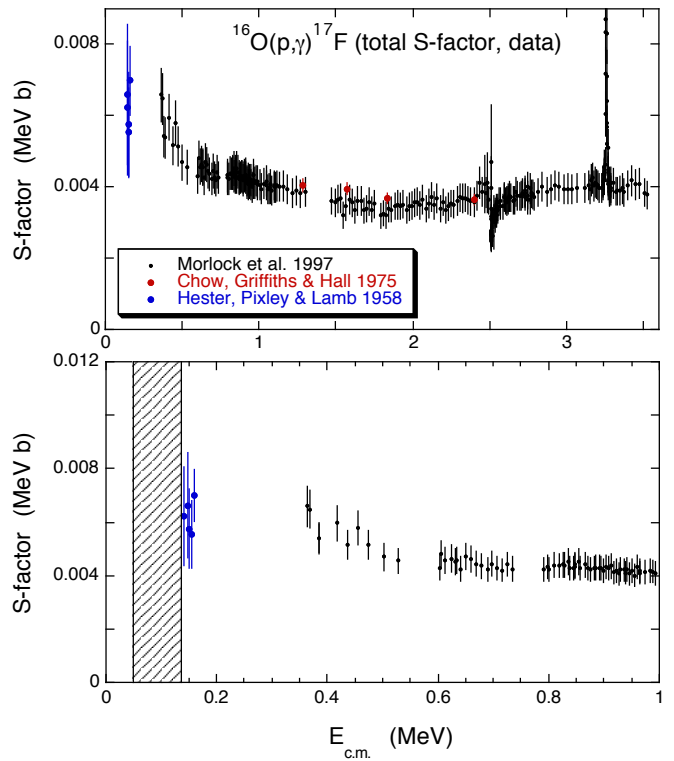}
\caption{\label{fig:totalsfactordata} 
(Color online) Total astrophysical S-factor for
the $^{16}$O(p,$\gamma$)$^{17}$F reaction versus center-of-mass
bombarding energy; (top) all data considered, (bottom) expanded view for low
energies. A number of different corrections are applied to the data of
Refs. \cite{HES58,MOR97,CHO75} while the results of
Refs. \cite{ROL73,TAN59} are disregarded in the present work; see text
for details. Note that we neglected the lowest energy data point of
Ref. \cite{HES58} because the associated error is very large
($\pm$52\%). The total S-factor varies slowly, except near the
vicinity of the lowest lying resonances at E$_p$=2.66 and 3.47 MeV
\cite{TIL93}. The shaded area in the lower part indicates the
effective energy range of stellar burning in massive AGB stars corresponding to
temperatures of T=60-100 MK.}
\end{figure}

\section{Nuclear reaction model calculations \label{nrmc}}

In this section we describe the fitting of a number of models to the data reported in Chow, Griffiths and Hall \cite{CHO75} and Morlock et al. \cite{MOR97}. The fits are performed for each transition separately. No attempt has been made in the
present work to constrain the fitting by using the reported total
S-factors of Hester, Pixley and Lamb \cite{HES58}. However, the latter
data, together with the single data point of Becker et
al. \cite{BEC82} for the transition to the first excited $^{17}$F
state, provide important cross checks for the present results. One of
our main goals is to extrapolate the S-factor to center-of-mass
energies of E$\approx$50--140 keV, corresponding to stellar
temperatures of T$\approx$0.06--0.1 GK that are characteristic of hot
bottom burning in AGB stars (see the shaded area in Fig. \ref{fig:totalsfactordata}). 
It will be seen that our procedures
give rise to significantly improved thermonuclear reaction rates over the entire
temperature range of astrophysical interest (T=0.01--10 GK).
Throughout this theoretical section and the following sections, all
kinematic quantities are given in the center--of--mass system, unless
mentioned otherwise.

A proper analysis should distinguish between statistical and systematic errors. We corrected the data of Refs. \cite{CHO75,MOR97} for a number of different effects
(Sec. \ref{dta}). However, the original papers do not provide sufficient information to quantitatively disentangle statistical and systematic errors. In view of this ambiguity, we decided to treat the combined uncertainties as statistical errors in the fitting procedure. The same problem and strategy is encountered in Ref. \cite{DAA04}. More information regarding our error analysis, and a justification for the adopted procedure, is given in Sec. \ref{reccS}.

The radiative capture cross section for $\opg$ from an initial state $J_i\pi_i$ to a final state $J_f\pi_f$ is given by
\beq
\sigma_c(J_i\pi_i \rightarrow J_f \pi_f)=\frac{\pi}{2k^2}(2J_i+1)
|U^{\gamma}(J_i\pi_i \rightarrow J_f \pi_f)|^2,
\label{def_sig}
\eeq
where $U^{\gamma}$ is the transition matrix element. For a given final state $J_f$, the total
cross section is obtained by summing over all initial angular momenta $J_i$.
This definition of the cross section
is common to all models; the only difference is the way of calculating the transition matrix element.
In the potential model, this term is computed from an integral involving the initial and final
wave functions, and the radial part of the electromagnetic operator. In the $R$-matrix approach, it
is split in two parts, involving different parameters fitted to experiment. We give more detail below on both models.

\subsection{Potential model}
We start the analysis of existing data by using a simple model for
the low--energy cross section in $^{16}$O(p,$\gamma$)$^{17}$F. The
model is referred to as ``direct proton capture" and it assumes a
single--step process where the proton is directly captured, without
formation of a compound nucleus, into a final bound state with the
emission of a photon. Many nonresonant reactions in nuclear
astrophysics have been described by this model (see Ref. \cite{ILI01}
and Tab. I in Ref. \cite{ILI04}).

Numerous studies (see, for example, Ref. \cite{ILI04}) have shown that
at relatively low bombarding energies and small binding energies of
the final states the direct capture of protons occurs mainly far
outside the nuclear radius. For this reason, the direct capture
process is sometimes referred to as ``extra--nuclear" capture. This
also implies that the calculated cross section is expected to be
relatively insensitive to the details of the model used to describe
the nuclear interior.

%
In the potential-model formalism the transition matrix element is given by
\begin{eqnarray}
 &&   U^{\gamma}(J_i\pi_i \rightarrow J_f \pi_f)  \sim   
    \sqrt{C^{2}\mathcal{S}} \nonumber \\
  &&\times     \int_{0}^{\infty} u_{i}(r,E) \, \mathcal{O}_{\omega \lambda}(r) \, u_{f}(r) 
         \, r^{2} \, dr 
         	 \label{eq:overlap}
\end{eqnarray}
where $C^{2}\mathcal{S}$ denotes the spectroscopic factor of the final state \cite{COM01}, and $\mathcal{O}_{\omega \lambda}$ is the radial part of the multipole operator for electromagnetic radiation of character $\omega$ (either electric
or magnetic) and multipolarity $\lambda$; $u_{i}$ denotes the radial part of the partial waves of the initial scattering state and $u_{f}$ is the radial wave function of final bound state. The model adopted here is similar to the one employed in Ref. \cite{MOR97}. The full formalism is given in Refs. \cite{Kim87,Mohr93}.

For example, for
the $^{16}$O(p,$\gamma$)$^{17}$F reaction the transition to the
ground state proceeds predominantly via E1 radiation and angular momenta of 
$\ell_i=1,3~(J_i=3/2^-,5/2^-,7/2^-) \rightarrow \ell_f=2~(J_f=5/2^+)$ (DC$\rightarrow$0), while the transition to the
first excited state at E$_x$=495 keV proceeds via E1 radiation and
angular momenta of $\ell_i=1~(J_i=1/2^-,3/2^-) \rightarrow\ell_f=0~
(J_f =1/2^+)$ (DC$\rightarrow$495). Other transitions, that is, those of M1 or E2 character, are negligible for the direct proton capture on
$^{16}$O. See also Ref.~\cite{DES99}.

The scattering and bound state wave functions are generated by an optical potential
\begin{equation}
V(r) = \xi V_F(r) + \xi_{s.o.} \frac{4}{r} \frac{dV_F}{dr} \vec L \cdot \vec S +V_C(r)
\label{eq:folding}
\end{equation}
where $V_F$ denotes a folding potential and $\xi$ is the folding potential strength parameter. Folding potentials have the major advantage that the geometry of the potential is fixed by the folding calculation. In other words, the potential can only be changed by a variation of the strength parameter $\xi$. The (weak) spin--orbit potential term has the usual Thomas form and is characterized by the spin--orbit strength parameter $\xi_{s.o.}$, while the Coulomb potential $V_C$ is given by a uniformly charged sphere of radius $R_{C}$. The folding potential is determined
%
%
using the effective nucleon--nucleon interaction adopted from the well--established DDM3Y parametrization. For the volume integral per interacting nucleon pair and the root--mean--square radius we find values of $J_R$=525.93 MeV fm$^3$ and $r_{F,rms}$=3.311 fm, respectively, while we also adopt the latter value for the Coulomb radius. For details regarding the folding procedure, see Refs. \cite{MOR97,ABE93}. 

For generating the bound state wave function, the parameter $\xi$ is adjusted to reproduce the binding energies of the ground and first excited state, while the parameter $\xi_{s.o.}$ can be constrained by the energy splitting of the lowest 5/2$^+$ and 3/2$^+$ states in $^{17}$F. We find values of $\xi$=1.0976 and $\xi_{s.o.}$=$-$0.1757 fm$^2$. For generating the scattering state wave function, it was shown in Ref. \cite{MOR97} that good agreement between the experimental capture data and the calculation is found by using a value of $\xi\approx$1.0 and the same value for $\xi_{s.o.}$ as for the bound state calculation. This potential also describes reliably the $^{16}$O(p,p)$^{16}$O elastic scattering data at low energies.

Once the folding potential strength parameter $\xi$ for generating the scattering wave function has been fixed, the spectroscopic factor $C^2\mathcal{S}$ is the only adjustable parameter in the above model. Our strategy will be as follows. Initially, the potential parameter is held constant at $\xi$=1.0 and the data of Chow, Griffiths and Hall \cite{CHO75} and Morlock et al. \cite{MOR97} are fitted independently. The least--square fits provide for each transition and data set the corresponding value of $C^2\mathcal{S}$ with an associated error. Afterward, the above procedure is repeated by systematically varying the value of $\xi$ in order to investigate the sensitivity of the extrapolated S-factor.

Numerous studies have shown that the spectroscopic factors for the
transitions to the ground and first excited states in $^{16}$O(p,$\gamma$)$^{17}$F are close to unity (see, for example, Ref. \cite{ILI04}). It must be emphasized, however, that we are here not at all concerned with the extraction of accurate spectroscopic factors from the measured low-energy cross sections. The magnitude of $C^2\mathcal{S}$ will depend on the potential parameter $\xi$, as is apparent from Eqs. \ref{eq:overlap} and \ref{eq:folding}. In
the present analysis, the spectroscopic factors are simply regarded as
intermediate results, or scaling factors, whose derived uncertainty
determines partially the error in the extrapolated astrophysical
S-factor.

\begin{figure}[]
\includegraphics[height=10cm]{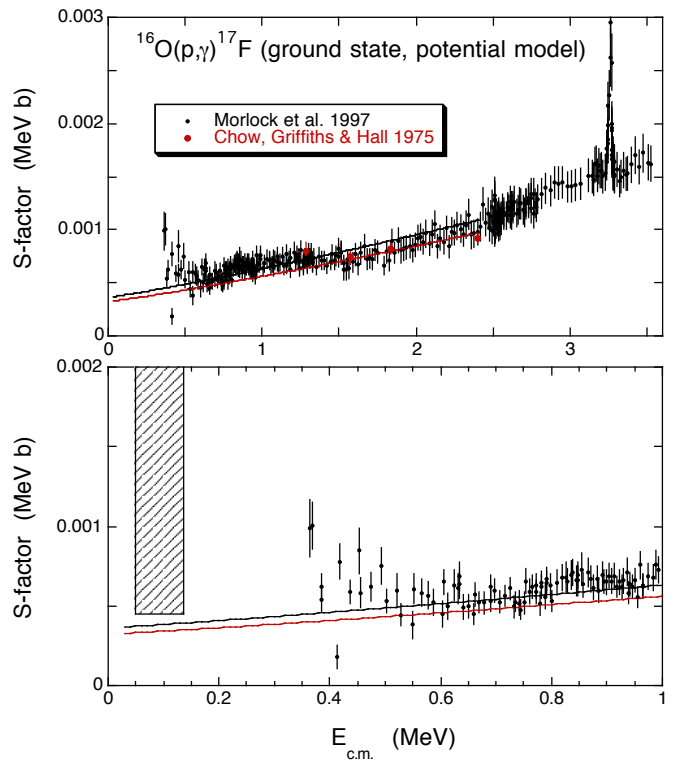}
\caption{\label{fig:groundsfactordata} 
(Color online) S--factor for the proton capture into the $^{17}$F ground state. The black and red lines represent fits to the data of Morlock et al. \cite{MOR97} and Chow, Griffiths and Hall \cite{CHO75}, respectively. The lines are obtained with a folding potential model (Sec. III.A). The 68\%--confidence bands are rather narrow and are not displayed for reasons of clarity. Only data below a center--of--mass energy of E=2.4 MeV are included in the fits. The shaded area in the lower part indicates the effective energy range for the hot bottom burning in AGB
stars.}
\end{figure}

\begin{figure}[]
\includegraphics[height=10cm]{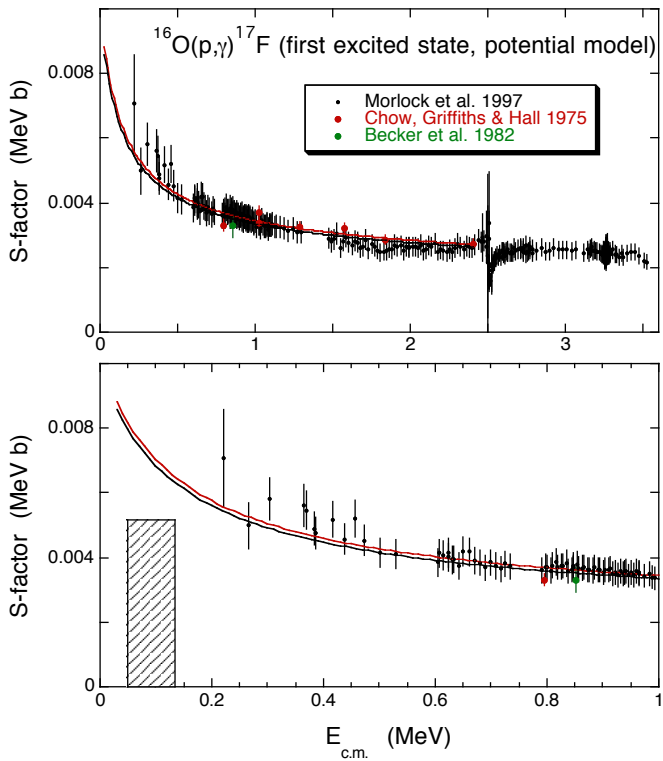}
\caption{\label{fig:firstsfactordata} 
(Color online) S--factor for the proton capture
into the first excited state of $^{17}$F. The black and red lines represent fits to the data of Morlock et al. \cite{MOR97} and Chow, Griffiths and Hall \cite{CHO75}, respectively. The lines are obtained with a folding potential model (Sec. III.A).
Only data below a center--of--mass energy of E=2.4 MeV are included in the fits. 
Notice the single data point reported by Becker et al. \cite{BEC82}, shown in green. See also caption of Fig. \ref{fig:groundsfactordata}.}
\end{figure}

Our potential model does not account for the presence of
resonances. Therefore, we only consider the center--of--mass energy
range below E=2.4 MeV for the least--squares fits described in this
section. The fits to the experimental S-factor data for the transition
to the $^{17}$F ground state are shown in Fig. \ref{fig:groundsfactordata}. The results are $C^2\mathcal{S}(DC\rightarrow 0)$=$1.09\pm$4.0\% ($\chi^2_{red}$=1.4) for the data of Chow, Griffiths and Hall \cite{CHO75} and
$C^2\mathcal{S}(DC\rightarrow 0)$=$1.22\pm$0.9\% ($\chi^2_{red}$=1.1)
for the data of Morlock et al. \cite{MOR97}. The absolute normalization of these two data sets differs thus by 12\%. Figure \ref{fig:firstsfactordata} shows the fits to the experimental S-factor data of the transition to the first excited state in $^{17}$F. In this case, we find $C^2\mathcal{S}(DC\rightarrow 495)$=$1.05\pm$1.7\%
($\chi^2_{red}$=1.5) for the data of Ref. \cite{CHO75} and $C^2\mathcal{S}(DC\rightarrow 495)$=$1.02\pm$0.5\% ($\chi^2_{red}$=0.3) for the data of Ref. \cite{MOR97}. The difference in the absolute normalization of the two data sets amounts to 3\%. Note that the upturn in the S-factor at low energies is well understood. See, for example, Refs. \cite{CHO75,MOR97}.

A different choice for the parameter $\xi$ of the scattering potential changes not only the magnitude, but also the energy--dependence, of the calculated direct capture S-factor. The sensitivity of the extrapolated S-factor to this parameter will be explored at a center--of--mass energy of E=90 keV which is located at the center of the energy region important for the hot bottom burning in AGB stars
(shaded area in Figs. \ref{fig:groundsfactordata} and \ref{fig:firstsfactordata}.). We quote in the following the results obtained for the dominant transition to the first excited state in $^{17}$F (DC$\rightarrow$ 495). For this sensitivity study we use the corrected experimental data of Ref. \cite{MOR97} with their statistical errors only (that is, excluding the additional 10\% overall uncertainty; see Sec.~II.B). Similar results are obtained for the data of Ref. \cite{CHO75}. For the reference calculation (using $\xi$=1) we find a spectroscopic factor of $C^2\mathcal{S} = 1.02$ and an S--factor of $S_{\rm{ref}}$(90 keV)= 7.03\,keV\,b for the DC$\rightarrow$495 transition. For example, a variation in the potential strength parameter $\xi$ by 10\% changes both the spectroscopic factor obtained from a least--squares fit of the data of Morlock et al. and the extrapolated S-factor at E=90 keV by about 5\%.  
The results of our sensitivity study for center--of--mass energies of $E$=90, 500 and 1000 keV are given in Tab. \ref{tab:ratio}. We also find that calculations with Woods-Saxon scattering potentials give similar results, even when the Woods-Saxon parameters are varied over a relatively broad range.

\begin{table}
  \caption{\label{tab:ratio} 
    Astrophysical S--factors for $^{16}$O(p,$\gamma$)$^{17}$F 
    at energies of $E = 90$, 500, and 1000\,keV.
    The results shown here are obtained by fitting the data of Ref. \cite{MOR97} for the dominant transition to the first excited state only. The S--factor obtained in the
    folding potential calculation of Ref. \cite{MOR97} is used as a
    reference, $S_{\rm{ref}}$. The parameter $\xi$ refers to the strength of the scattering potential. 
}
\begin{center}
\begin{tabular}{ccccc}\hline\hline
E (keV) & 90 &  500 & 1000 & \\
$S_{\rm{ref}}(E)$ (keV\,b) & 7.03 & 4.20 & 3.39 & \\
$S_{\rm{ref}}(E)/S_{\rm{ref}}(1\,{\rm{MeV})}$ & 2.09 & 1.24 & $\equiv 1.00$ & \\
\hline
$\xi$ & \multicolumn{3}{c}{$S(E)/S_{\rm{ref}}(E)$} & $\chi^2_{red}$\\
\hline
$0.8$ & 1.03 & 1.07 & 1.01 & 1.2 \\
$0.9$ & 1.05 & 1.03 & 1.01 & 1.1 \\
$1.0$ & $\equiv 1.00$ & $\equiv 1.00$ & $\equiv 1.00$ & 2.2 \\
$1.1$ & 0.96 & 0.97 & 0.99 & 4.7 \\
%
\hline
\hline
\end{tabular}
\end{center}
\end{table}

It is interesting to note that a variation in the potential strength parameter $\xi$ influences the calculated energy dependence of the S--factor which, in turn, gives rise to a noticeable variation in the $\chi^2_{red}$ value for the
adjustment of the spectroscopic factor $C^2\mathcal{S}$. A prerequisite for such a study is a large number of data points as provided by Ref. \cite{MOR97}. For the folding potential we find a broad $\chi^2$ minimum near $\xi \approx 0.9$. The value of $\chi^2_{\rm{red}}$ increases by 1 if $\xi$ is changed by about 10\,\%. Thus, the
energy dependence of the S--factor confines the potential strength parameter to the range explored in Tab.~\ref{tab:ratio}. 

In summary, a variation of the potential parameter $\xi$ within a
reasonable physical range (10\%) changes the extrapolated S-factor by 
$\approx$1--5\%, depending on the energy $E$. This value can be compared to the statistical error obtained from the least--squares fits alone (0.5--4\%) and to the difference between the absolute normalizations of the two data sets under consideration (3-12\%).


\begin{figure}[]
\includegraphics[height=10.5cm]{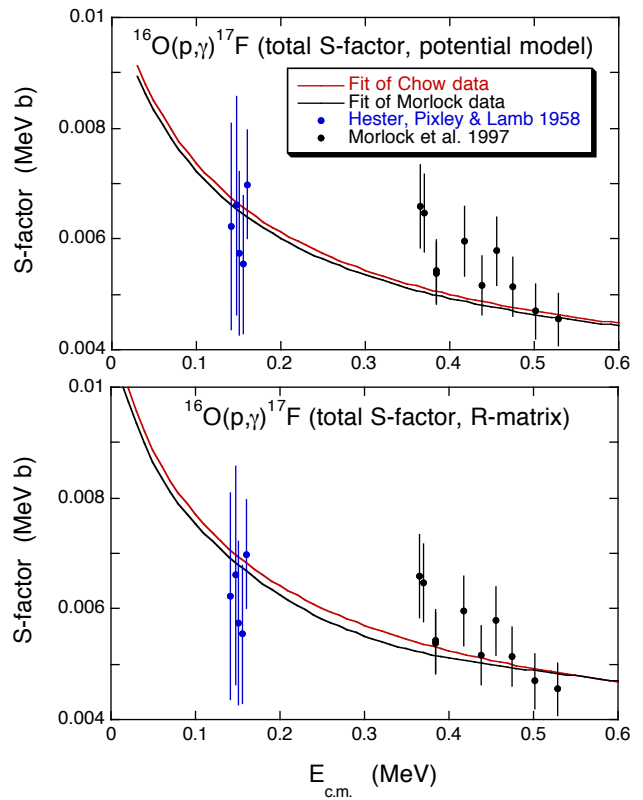}
\caption{\label{fig:totalsfactordata_low} 
(Color online) Total S--factor for $^{16}$O(p,$\gamma$)$^{17}$F at low bombarding energies. The red and black lines indicate fits to the data of Chow, Griffiths and Hall \cite{CHO75} and Morlock et al. \cite{MOR97}, respectively, by using a folding potential model (top part; see Sec. III.A) and an R-matrix model (bottom part; see Sec. III.B). The low--energy data of Hester, Pixley and Lamb
\cite{HES58} are shown for comparison.}
\end{figure}

The sum of the S-factors for both transitions is shown in
Fig. \ref{fig:totalsfactordata_low} (top). The red and black lines indicate the
fits to the data of Chow, Griffiths and Hall \cite{CHO75} and Morlock et al. \cite{MOR97},
respectively. It can be seen that the analyses of both data sets give
consistent results. As a cross--check, the low--energy data of Hester,
Pixley and Lamb \cite{HES58} are also shown. Recall that these authors
only report the total S--factor. Clearly, the solid lines are
consistent with the data of Ref. \cite{HES58}, providing further
support to our results.

\subsection{R-matrix model}
\subsubsection{Formalism}
The R-matrix theory combines the flexibility of a phenomenological approach with the physical content of the Coulomb wave functions. It can be applied to resonant as well as to non-resonant reactions. In the R-matrix theory the relative coordinate $r$ between the colliding nuclei is 
divided into two sectors, limited by the channel radius $a$. In the internal region ($r\leq a$) the
physics of the problem is determined from the properties of poles. In the external region
($r > a$), the relative wave function is given by a linear combination of Coulomb
wave functions. This procedure provides an accurate extrapolation down to low energies,
where Coulomb effects are expected to play a major role. The poles are associated with
resonances or bound states of the unified nucleus, and their properties (energy, particle
and $\gamma$-ray partial widths) are fitted to the available data.

The formalism of the R-matrix method is well known \cite{LT58}. Here we use the formulation
of Ref.~\cite{DAA04}. 
The electromagnetic transition matrix element $U^{\gamma}$ involved in Eq.~(\ref{def_sig})
%
%
is divided into two parts, associated with the internal and external regions, as
\beq
U^{\gamma}=U^{\gamma}_{int}+U^{\gamma}_{ext}
\label{def_u}
\eeq
We present the different contributions in Ref.~\cite{DAA04}. The internal part $U^{\gamma}_{int}$ is given by the energies and partial widths of the poles, while the external contribution $U^{\gamma}_{ext}$ is provided by an integral involving Coulomb wave functions and is evaluated from the channel radius $a$ to infinity. This external term is proportional to the asymptotic normalization constant (ANC) of the final (bound) state. Let us point out that both contributions are related to each other, as the external term involves the phase shift which, in turn, depends on the R matrix.

The R-matrix theory can be easily extended to non-resonant reactions. In that case one usually assumes that the non-resonant contribution is simulated by a high-energy background pole. Although this method is acceptable, it presents some drawbacks: the results are somewhat sensitive to the choice of the pole energy, which in addition has no genuine physical meaning. This problem can be addressed by assuming that the R-matrix is constant. Non-resonant transfer reactions have been investigated previously in that way \cite{AD98}. The R-matrix is therefore given by
\beq
R(E)=R^p_0
\eeq
where subscript $p$ refers to the proton channel. When $R^p_0=0$, the problem is equivalent to the hard-sphere approximation. 

This procedure can be extended to capture reactions. Let us consider the internal contribution in
Eq.~(\ref{def_u}). According to Ref.~\cite{DAA04}, we have in a general case
\beq
U^{\gamma}_{int}=i^{\ell}\exp(i\delta_{\rm HS})  \frac{1}{1-LR} 
\sum_{i=1}^N \frac{ \sqrt{\tilde{\Gamma}_i \tilde{\Gamma}_{\gamma,i}}}{E_i-E}
\label{u_int}
\eeq
where $N$ is the number of poles, $(E_i,\tilde{\Gamma}_i,\tilde{\Gamma}_{\gamma,i})$ are their
energies, particle and $\gamma$-ray partial widths, $\delta_{\rm HS}$ is the hard-sphere phase shift, and the constant $L$ is related to the Coulomb wave functions (see Refs.~\cite{LT58,DAA04} for details). Including the energy dependence of the widths we have
\beq
\tilde{\Gamma}_i \tilde{\Gamma}_{\gamma,i}=2P_{\ell}(E)\tilde{\gamma}_i^2 \tilde{\Gamma}^0_{\gamma,i}
(E-E_f)^{2\lambda+1}
\label{gaga}
\eeq
where $E_f$ is the energy of the final state, $\lambda$ is the order of the multipole, and 
$\tilde{\gamma}_i^2, \tilde{\Gamma}^0_{\gamma,i}$ are energy-independent quantities. For a non-resonant reaction ($N=1$), Eq.~(\ref{u_int}) can be rewritten as
\begin{widetext}
\beq
U^{\gamma}_{int}=i^{\ell}\exp(i\delta_{\rm HS})\frac{1}{1-LR^p_0}
\bigl[ 2R^p_0 R^{\gamma}_0 P_{\ell_i}(E)(E-E_f)^{2\lambda+1}\bigr]^{1/2}
\eeq
\end{widetext}
where $R^{\gamma}_0$ is a constant connected with the electromagnetic transition. This parametrization is equivalent to the usual ``background" term, but without the redundancy associated with the energy of the background pole. Note that a non-resonant capture transition is characterized by the fitting parameters $R^p_0$, $R^{\gamma}_0$ and the ANC of the final state.

\subsubsection{Results}
For the ground-state transition in $\opg$, we have $\ell_i=1\ (J_i=3/2^-$), and a small contribution from $\ell_i=3\ (J_i=5/2^-,7/2^-$). All these transitions are non-resonant. For the excited-state contribution $(J_f=1/2^+)$, we include the $1/2^-$ resonance at $E_r=2.50$ MeV, with parameters taken from Ref.~\cite{TIL93}. The $J_i=3/2^-$ partial wave is also included as a non-resonant contribution.  For simplicity we assume that the partial waves of the same spin-orbit
doublet (that is, $J_i=5/2^-$ and $J_i=7/2^-$) have the same R-matrix parameters. For the channel radius we used a value of $a$=5~fm. A simultaneous fit for the ground and excited states is then performed.

\begin{figure}[]
\includegraphics[height=10cm]{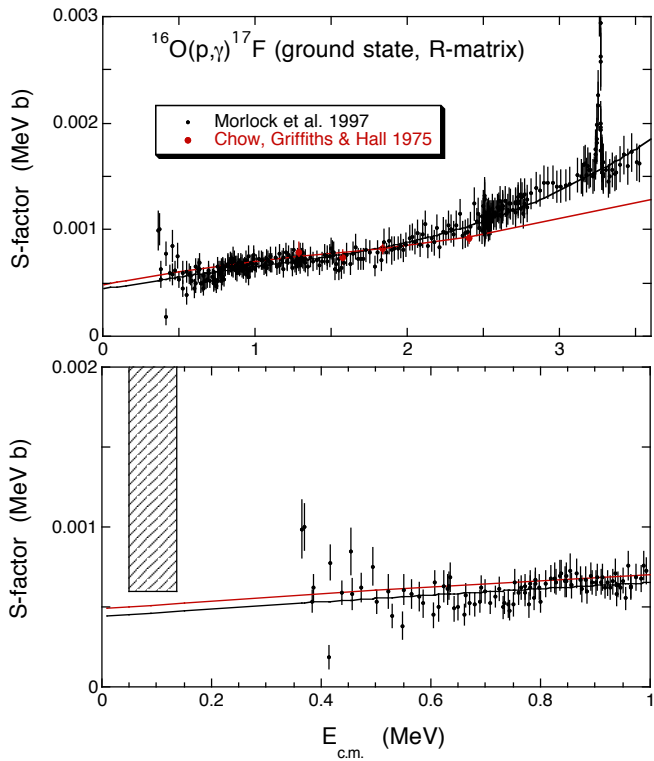}
\caption{\label{fig:groundsfactordataR} 
(Color online) S--factor for the proton capture into the $^{17}$F ground state. The black and red lines represent fits to the data of Morlock et al. \cite{MOR97} and Chow, Griffiths and Hall \cite{CHO75}, respectively. The lines are obtained with an R-matrix model (Sec. III.B). The shaded area in the lower part indicates the effective energy range for the hot bottom burning in AGB stars.}
\end{figure}

\begin{figure}[]
\includegraphics[height=10cm]{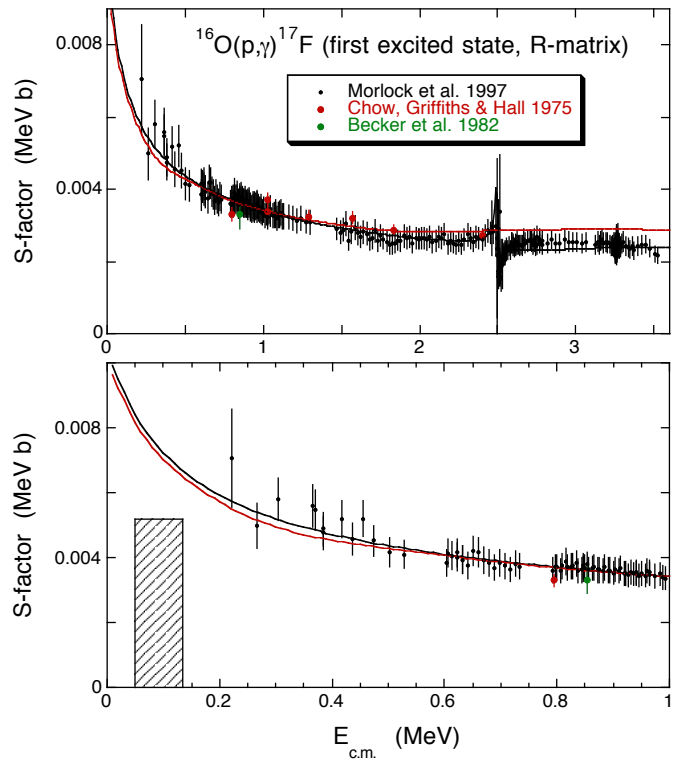}
\caption{\label{fig:firstsfactordataR} 
(Color online) S--factor for the proton capture into the first excited state of $^{17}$F. The black and red lines represent fits to the data of Morlock et al. \cite{MOR97} and Chow, Griffiths and Hall \cite{CHO75}, respectively. The lines are obtained with an R-matrix model (Sec. III.B). The shaded area in the lower part indicates the effective energy range for the hot bottom burning in AGB stars.}
\end{figure}

The R-matrix fits for the transitions to the $^{17}$F ground and first excited state are shown in
Figs.  \ref{fig:groundsfactordataR} and \ref{fig:firstsfactordataR}, respectively. For the asymptotic normalization constant of the ground state transition we find values of ANC=1.19$\pm$0.02 fm$^{-1/2}$ ($\chi^2_{red}$=0.2) for the data of Chow, Griffiths and Hall \cite{CHO75} and ANC=1.13$\pm$0.01 fm$^{-1/2}$ ($\chi^2_{red}$=0.6) for the data of Morlock et al. \cite{MOR97}. The absolute normalization of these two data sets differs by 5\%. For the transition to the first excited $^{17}$F state, the results are ANC=81.0$\pm$0.9 fm$^{-1/2}$ ($\chi^2_{red}$=1.6) for the data of Ref. \cite{CHO75} and ANC=82.3$\pm$0.3 fm$^{-1/2}$ ($\chi^2_{red}$=0.7) for the data of Ref. \cite{MOR97}, where the absolute normalization differs by 2\%. Note that the asymptotic normalization constants from our R-matrix analysis are consistent with the microscopic results of Ref.~\cite{BDH98}. 

The sum of the S-factors for both transitions is displayed in Fig. \ref{fig:totalsfactordata_low} (bottom). The red and black lines show the results from fitting the data of Chow, Griffiths and Hall \cite{CHO75} and Morlock et al. \cite{MOR97}, respectively. It is apparent that the R-matrix analyses of both data sets give consistent results and describe the total S-factor data rather well. 
The channel radius is adopted here as $a=5$ fm, a typical value used in the literature \cite{DAA04}.
The sensitivity of the extrapolated S--factor to systematic variations of the channel radius $a$ were investigated at a center-of-mass energy of 90 keV. We find that a variation of the channel radius by 10\% changes the extrapolated S--factor at E=90 keV by about 2\%. 


\section{Recommended S-factor and reaction rates}\label{reccS}

In the previous sections we presented the total S--factors and their components for two different data sets, analyzed by using two different nuclear reaction models. We will now describe our method of estimating recommended S--factors and their associated errors. This information will then be used in order to calculate new recommended reaction rates.

\begin{figure}[]
\includegraphics[height=6.5cm]{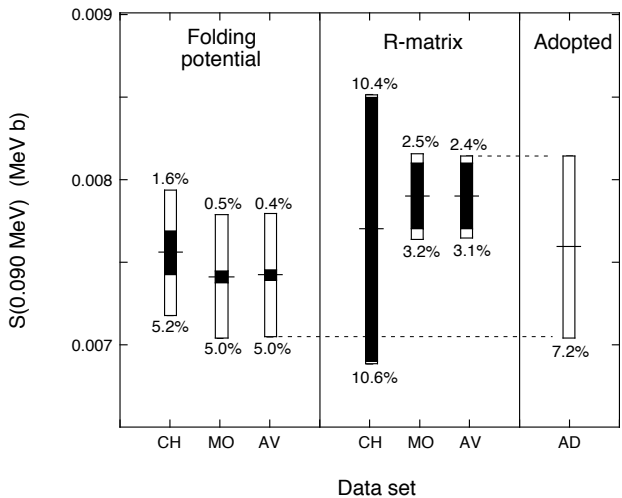}
\caption{\label{fig:compresults} 
Total S--factor for $^{16}$O(p,$\gamma$)$^{17}$F at a center--of-mass bombarding energy of E=0.090 MeV, shown here to illustrate the estimation of the recommended S--factor. For an explanation of the x--axis labels and other details, see text.}
\end{figure}

The method is explained in Fig. \ref{fig:compresults} showing, as an example, the total 
S--factor at an energy of E=0.090 MeV, that is, at the center of the effective energy range 
important for the hot bottom burning in AGB stars. The left and middle panels present 
results obtained from the folding potential model and the R-matrix approach, respectively. 
The x--axis labels ``CH" and ``MO" refer to the analysis of the data of Ref. \cite{CHO75} 
and Ref. \cite{MOR97}, respectively. The solid bars indicate 1$\sigma$ errors resulting from 
the fitting of the data alone (statistical errors). Note that in general the errors obtained 
in the R-matrix approach with three fitting parameters in our case are larger than those 
derived from the potential model with just one fitting parameter. The errors indicated by 
the open bars are obtained if systematic model uncertainties (i.e., those obtained by varying the 
parameters of the model within a reasonable physical range) are added 
quadratically to the statistical errors. The numerical values of the statistical and total 
errors (in percent) are listed at the top and bottom, respectively, of each error bar. The 
label ``AV" refers, for each model, to the weighted average of the two data sets, where only 
statistical errors are used for the weighing procedure and systematic model errors are added 
quadratically afterwards. The right panel displays the final adopted S-factor (label ``AD"). 
Its total error is given by the extrema of the average values resulting from the two 
reaction models. The mean value of the final recommended S-factor is then given by the 
arithmetic average of the upper and lower bound of the error bar (open bar in right panel).

It was pointed out in Sec. \ref{mormor} that an additional uncertainty of 10\% has to be taken into account for the data of Morlock et al. Unfortunately, insufficient information is provided in Ref. \cite{MOR97} to determine how much of this uncertainty arises from systematic and from statistical effects. Recall that in the above analysis we treat the entire additional uncertainty as a statistical error in the fitting procedure (see Sec. \ref{nrmc}). Our treatment will tend to underestimate the true error. Alternatively, one may regard the entire additional uncertainty as a systematic error, which would tend to overestimate the true error. Clearly, the true error must be located somewhere between the two extremes. Test calculations have been performed in order to quantify this effect. Consider as an example the folding potential analysis of the dominating transition to the first excited state for an energy of E=0.090 MeV (similar to the left-hand side of Fig. 8). The alternative assumption of treating the additional uncertainty as a systematic error would increase the averaged (``AV") S-factor by 2.6\%, while its uncertainty would increase from 5.1\% to 5.3\%. Similar results are obtained in the R-matrix analysis. It is obvious that such small variations are negligible for the adopted S-factor (``AD").

The thermonuclear rate per particle pair for a reaction involving two nuclei is given by \cite{ILI07}
\begin{equation}  
N_A \langle \sigma v \rangle = N_A \frac{\sqrt{8/(\pi m)}}{(kT)^{3/2}} \int_0^\infty S(E)\,e^{-E/kT-2\pi\eta}\,dE \label{reactionratepp}
\end{equation}  
where $S(E)$ is the S-factor at energy $E$ and the factor $e^{-E/kT}$ derives from the Maxwell-Boltzmann distribution. The quantities $T$, $k$, $N_A$ and $m$ are the plasma temperature, the Boltzmann constant, Avogadro's constant and the reduced mass of the interacting nuclei, respectively. The thermonuclear rates for the $^{16}$O(p,$\gamma$)$^{17}$F reaction, calculated by integrating Eq.~\ref{reactionratepp} numerically using the $S$--factor recommended in this work, are listed in Tab.~\ref{tab:reactionrates} for temperatures in the range of T=0.01-2.5 GK. For higher temperatures, the reaction rates are influenced by energies not covered in the present work (E$>$2.4 MeV).

\begin{figure}[]
\includegraphics[height=6.5cm]{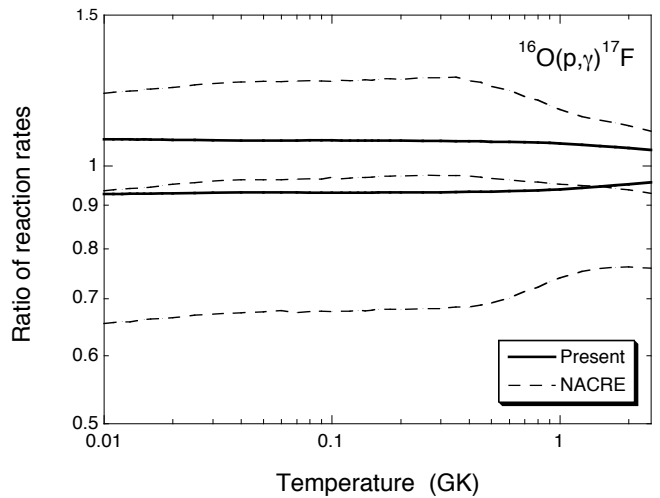}
\caption{\label{fig:rates} 
Reaction rates of $^{16}$O(p,$\gamma$)$^{17}$F. For better comparison, we show the lower bound, recommended rate and upper bound on the rate normalized to the present recommended rate. Solid and dashed lines indicate the rate ratios for the present and the NACRE \cite{ANG99} results, respectively.}
\end{figure}

Results for the reaction rates of $^{16}$O(p,$\gamma$)$^{17}$F are shown in 
Fig.~\ref{fig:rates}. The curves display the ratios of lower bound, recommended rate and upper bound on the rate over the present recommended rate. Solid and dashed lines indicate the rate ratios resulting from our work and from NACRE \cite{ANG99}, respectively. It can be seen that the ratio of the NACRE and the present {\it recommended} rates is close to unity. However, below a temperature of T=0.5 GK, the present reaction rate errors amount to only $\approx$7\%. This represents an improvement by a factor of $\approx$4 over the results published by NACRE which followed a rather conservative rate evaluation procedure. Note that because of  extensive efforts, the $^{16}$O(p,$\gamma$)$^{17}$F reaction has now the most precisely known rate involving any target nucleus in the mass $A\geq12$ range. Astrophysical consequences for the oxygen isotopic ratios from hot bottom burning in AGB stars are presented in the following section.

\begin{table}[tbh]
\caption{\label{tab:reactionrates}Total thermonuclear reaction rates for $^{16}$O(p,$\gamma$)$^{17}$F in units of (cm$^3$mol$^{-1}$s$^{-1}$).
}
\begin{ruledtabular}
\begin{tabular}{ c c c c }
T (GK)    &  Low    & Recommended   & High  \\
\hline 
0.01 		& 6.674$\times$10$^{-25}$  	& 7.200$\times$10$^{-25}$  	&  7.733$\times$10$^{-25}$  \\
0.011 	& 7.026$\times$10$^{-24}$  	& 7.578$\times$10$^{-24}$  	&  8.138$\times$10$^{-24}$  \\
0.012 	& 5.638$\times$10$^{-23}$  	& 6.080$\times$10$^{-23}$  	&  6.528$\times$10$^{-23}$  \\
0.013 	& 3.626$\times$10$^{-22}$  	& 3.910$\times$10$^{-22}$  	&  4.197$\times$10$^{-22}$  \\
0.014 	& 1.941$\times$10$^{-21}$  	& 2.093$\times$10$^{-21}$  	&  2.246$\times$10$^{-21}$  \\
0.015 	& 8.910$\times$10$^{-21}$  	& 9.604$\times$10$^{-21}$ 	&  1.031$\times$10$^{-20}$  \\
0.016 	& 3.587$\times$10$^{-20}$  	& 3.866$\times$10$^{-20}$  	&  4.148$\times$10$^{-20}$  \\
0.018 	& 4.211$\times$10$^{-19}$  	& 4.536$\times$10$^{-19}$ 	&  4.866$\times$10$^{-19}$  \\
0.02 		& 3.505$\times$10$^{-18}$  	& 3.775$\times$10$^{-18}$  	&  4.048$\times$10$^{-18}$  \\
0.025 	& 2.431$\times$10$^{-16}$  	& 2.616$\times$10$^{-16}$  	&  2.803$\times$10$^{-16}$  \\
0.03 		& 6.124$\times$10$^{-15}$ 	& 6.586$\times$10$^{-15}$  	&  7.053$\times$10$^{-15}$  \\
0.04 		& 6.659$\times$10$^{-13}$  	& 7.155$\times$10$^{-13}$  	&  7.657$\times$10$^{-13}$  \\
0.05 		& 1.847$\times$10$^{-11}$  	& 1.984$\times$10$^{-11}$  	&  2.123$\times$10$^{-11}$  \\
0.06 		& 2.309$\times$10$^{-10}$  	& 2.481$\times$10$^{-10}$  	&  2.655$\times$10$^{-10}$  \\
0.07 		& 1.726$\times$10$^{-09}$  	& 1.855$\times$10$^{-09}$  	&  1.985$\times$10$^{-09}$  \\
0.08 		& 9.029$\times$10$^{-09}$  	& 9.706$\times$10$^{-09}$  	&  1.039$\times$10$^{-08}$  \\
0.09 		& 3.644$\times$10$^{-08}$  	& 3.917$\times$10$^{-08}$  	&  4.193$\times$10$^{-08}$  \\
0.1 		& 1.208$\times$10$^{-07}$  	& 1.299$\times$10$^{-07}$  	&  1.391$\times$10$^{-07}$  \\
0.11 		& 3.439$\times$10$^{-07}$  	& 3.697$\times$10$^{-07}$  	&  3.957$\times$10$^{-07}$  \\
0.12 		& 8.662$\times$10$^{-07}$  	& 9.312$\times$10$^{-07}$  	&  9.966$\times$10$^{-07}$  \\
0.13 		& 1.975$\times$10$^{-06}$  	& 2.124$\times$10$^{-06}$  	&  2.273$\times$10$^{-06}$  \\
0.14 		& 4.149$\times$10$^{-06}$  	& 4.460$\times$10$^{-06}$  	&  4.773$\times$10$^{-06}$  \\
0.15 		& 8.133$\times$10$^{-06}$  	& 8.742$\times$10$^{-06}$  	&  9.355$\times$10$^{-06}$  \\
0.16 		& 1.504$\times$10$^{-05}$  	& 1.616$\times$10$^{-05}$  	&  1.729$\times$10$^{-05}$  \\
0.18 		& 4.449$\times$10$^{-05}$ 	& 4.781$\times$10$^{-05}$  	&  5.114$\times$10$^{-05}$  \\
0.2 		& 1.129$\times$10$^{-04}$  	& 1.213$\times$10$^{-04}$  	&  1.298$\times$10$^{-04}$  \\
0.25 		& 7.231$\times$10$^{-04}$  	& 7.766$\times$10$^{-04}$  	&  8.305$\times$10$^{-04}$  \\
0.3 		& 2.954$\times$10$^{-03}$  	& 3.172$\times$10$^{-03}$  	&  3.391$\times$10$^{-03}$  \\
0.35 		& 9.035$\times$10$^{-03}$  	& 9.696$\times$10$^{-03}$  	&  1.036$\times$10$^{-02}$  \\
0.4 		& 2.262$\times$10$^{-02}$  	& 2.426$\times$10$^{-02}$  	&  2.592$\times$10$^{-02}$  \\\
0.45 		& 4.896$\times$10$^{-02}$  	& 5.250$\times$10$^{-02}$  	&  5.605$\times$10$^{-02}$  \\
0.5 		& 9.495$\times$10$^{-02}$  	& 1.018$\times$10$^{-01}$  	&  1.086$\times$10$^{-01}$  \\
0.6 		& 2.813$\times$10$^{-01}$  	& 3.012$\times$10$^{-01}$  	&  3.212$\times$10$^{-01}$  \\
0.7 		& 6.659$\times$10$^{-01}$  	& 7.122$\times$10$^{-01}$  	&  7.587$\times$10$^{-01}$  \\
0.8 		& 1.350$\times$10$^{+00}$ 	& 1.442$\times$10$^{+00}$  	&  1.535$\times$10$^{+00}$  \\
0.9 		& 2.447$\times$10$^{+00}$  	& 2.610$\times$10$^{+00}$  	&  2.775$\times$10$^{+00}$  \\
1.0 		& 4.074$\times$10$^{+00}$  	& 4.340$\times$10$^{+00}$  	&  4.607$\times$10$^{+00}$  \\
1.25 		& 1.123$\times$10$^{+01}$  	& 1.192$\times$10$^{+01}$  	&  1.261$\times$10$^{+01}$  \\
1.5 		& 2.418$\times$10$^{+01}$  	& 2.557$\times$10$^{+01}$  	&  2.696$\times$10$^{+01}$  \\
1.75 		& 4.437$\times$10$^{+01}$  	& 4.677$\times$10$^{+01}$  	&  4.918$\times$10$^{+01}$  \\
2.0 		& 7.295$\times$10$^{+01}$  	& 7.667$\times$10$^{+01}$  	&  8.041$\times$10$^{+01}$  \\
2.5 		& 1.586$\times$10$^{+02}$  	& 1.659$\times$10$^{+02}$  	&  1.731$\times$10$^{+02}$  \\
\end{tabular}
\end{ruledtabular}
\end{table}

\section{Hot bottom burning in AGB stars}

The aim of this section is to test the effect of our new $^{16}$O(p,$\gamma$)$^{17}$F 
reaction rate on the $^{17}$O/$^{16}$O ratios produced by HBB in massive AGB stars. This is of relevance because of the presence of a very small fraction of stellar meteoritic 
oxide grains that might have originated in massive AGB stars. These stars may have 
contributed to the abundances of some radioactive nuclei in the early solar system, for example, $^{26}$Al and $^{60}$Fe, as well as to other abundance anomalies observed in meteoritic material, for example, $^{62}$Ni, $^{87}$Rb, and $^{96}$Zr \cite{trigo07}. It is important to verify if there is also a population of meteoritic stellar grains that may have originated from this stellar site. 

The few (in fact, four) grain candidates for a massive AGB origin exhibit number abundance ratios of $^{18}$O/$^{16}$O $< 10^{-4}$, $^{17}$O/$^{16}$O between $7 \times 10^{-4}$ and $1.5 \times 10^{-3}$, and large $^{26}$Mg excesses caused by the decay of $^{26}$Al, with inferred initial $^{26}$Al/$^{27}$Al ratios of $\simeq$ 0.01-0.09. One of them, spinel grain OC2, also shows an excess in $^{25}$Mg. Such signatures could be produced in massive AGB stars via the combined activation of HBB and the operation of the $^{22}$Ne$+\alpha$ reaction during unstable He burning (thermal pulses). For a detailed discussion, see Lugaro et al. \cite{LUG07}. We focus here on the 
$^{17}$O/$^{16}$O ratio, which is the most problematic signature to be interpreted as produced by HBB (see also Ref. \cite{boothroyd95}). 

During HBB the abundance of $^{17}$O quickly reaches equilibrium. Thus the $^{17}$O/$^{16}$O ratio is mostly determined by the ratio of the reaction rates that produce and destroy $^{17}$O, that is, the $^{16}$O(p,$\gamma$)$^{17}$F and $^{17}$O(p,$\alpha$)$^{14}$N reactions, respectively. Because the laboratory measurements of the composition of meteoritic stellar grains provide isotopic ratios of very high precision, with error bars on the order of a few percent only, small uncertainties in the reaction rates are indispensable in this case to address the possible match of the models with the observations.

We consider here a stellar model of a typical massive (6.5 M$_{\odot}$) AGB star of solar metallicity ($Z=0.02$) experiencing HBB. This model was computed by Karakas \& Lattanzio \cite{karakas07} using mass loss on the AGB from Ref. \cite{vassiliadis93} and it experienced 39 thermal pulses. The temperature at the base of the convective envelope during the AGB phase increases from T=60 MK to T=87 MK at the 25$^{\rm th}$ thermal pulse, and then decreases again to T=20 MK at the end of the computed evolution. The high temperature reached by this model allows for the production of enough $^{26}$Al to match the composition of grain OC2, while the $^{17}$O/$^{16}$O ratio could be reproduced within the previous \cite{ANG99} $^{16}$O(p,$\gamma$)$^{17}$F reaction rate uncertainties (see Fig. 7 of \cite{LUG07}).

Similar to the procedure of Lugaro et al. \cite{LUG07}, we analyze the impact of the reaction rate uncertainties using a post-processing nucleosynthesis code. Hence we disregard any feedback of the different reaction rates on the stellar structure. This is justified in this case because: (i) the $^{16}$O(p,$\gamma$)$^{17}$F and $^{17}$O(p,$\alpha$)$^{14}$N reactions generate only a small amount of energy within the CNO cycle; and (ii) the $^{16}$O(p,$\gamma$)$^{17}$F reaction is only marginally activated in our model (only 23\% of the initial $^{16}$O is destroyed by the second dredge-up and by HBB). For the $^{16}$O(p,$\gamma$)$^{17}$F reaction we use our new rates, while for the $^{17}$O(p,$\alpha$)$^{14}$N reaction we employ the same rate as in Ref. \cite{LUG07} which is based on the latest available experimental information \cite{chafa05,fox05}. Different runs of the 6.5 M$_{\odot}$ stellar model are then computed by varying independently the rates of these two reactions between their lower and upper bounds. The results of this procedure are given in Tab.~\ref{tab:HBBresults}, showing how the errors in the rates propagate to uncertainties of the computed $^{17}$O/$^{16}$O ratio. Note that the calculated $^{17}$O/$^{16}$O ratio is almost the same as in Ref. \cite{LUG07} because the {\it recommended} $^{16}$O(p,$\gamma$)$^{17}$F reaction rate did not change significantly (see Fig.~\ref{fig:rates}). The main result here is that the total uncertainty in the derived abundance ratio is reduced from $+63$\% and $-73$\% when using the NACRE rate for $^{16}$O(p,$\gamma$)$^{17}$F to $+35$\% and $-30$\% when employing our new reaction rate.

\begin{table}
\caption{\label{tab:HBBresults} 
Final $^{17}$O/$^{16}$O abundance ratios for 6.5M$_{\odot}$, $Z=0.02$ stellar models computed with all possible combinations of $^{16}$O(p,$\gamma$)$^{17}$F and 
$^{17}$O(p,$\alpha$)$^{14}$N reaction rates: LR=lower bound of rate 
(at $T=60 - 100$ MK), RR=recommended rate, and UR=upper bound on rate (at $T=60 - 100$ MK). The $^{17}$O/$^{16}$O ratio is explicitly given for the RR-RR combination in the center of the table, while percent variations with respect to this case are given for the other rate combinations. Results for the $^{16}$O(p,$\gamma$)$^{17}$F rate derived in this paper are in roman font, 
results using the NACRE \cite{ANG99} rate are in italics.}
\begin{center}
\begin{tabular}{cc|ccc}
\hline\hline
 & & & $^{16}$O(p,$\gamma$)$^{17}$F & \\
 & & LR ($-$7\%, {\it $-$43\%}) & RR & UR ($+7$\%, {\it +30\%})\\
\hline
 & LR ($\simeq -$22\%) & $+$17\%, {\it $-$14\%} & $+$26\%, {\it $+$25\%} & $+$35\%, {\it $+$63\%} \\
$^{17}$O(p,$\alpha$)$^{14}$N & RR & $-$7\%, {\it $-$43\%} & 2.52$\times 10^{-3}$, {\it 2.44$\times {\it 
10}^{-{\it 3}}$} & 
$+$7\%, {\it $+$30\%} \\
 & UR ($\simeq +$18\%) & $-$30\%, {\it $-$73\%} & $-$20\%, {\it $-$21\%} & $-$13\%, {\it $+$7\%} \\
\hline
\hline
\end{tabular}
\end{center}
\end{table}

In conclusion, the measured $^{17}$O/$^{16}$O ratio of grain OC2 ($=1.25 \pm 0.07 \times 10^{-3}$) could be reproduced within the large error bars of the NACRE compilation ($2.44^{+1.54}_{-1.78}\times10^{-3}$) in models of massive AGB stars; however, the much more precise $^{16}$O(p,$\gamma$)$^{17}$F rate of the present work leads to $2.52^{+0.88}_{-0.76}\times10^{-3}$ for the $^{17}$O/$^{16}$O ratio and  disagrees with the measured value. Consequently, there is not clear evidence to date for any stellar grain origin from massive AGB stars. Stellar model uncertainties, such as different mixing prescriptions \cite{ventura05a} and mass loss rates \cite{ventura05b} still need to be carefully evaluated in this context. Another possibility is that we have not yet discovered grains from massive AGB stars because they are perhaps smaller in size than the grains currently analyzed in the laboratory ($>$~1~$\mu$m) \cite{bernatowicz05,nuth06}. Clearly, a solution to this problem requires future efforts.

\section{Summary}

In this work we focused our attention on a reanalysis of the $^{16}$O(p,$\gamma$)$^{17}$F reaction rates. We started from an evaluation of all the original data and performed a number of corrections when appropriate. The modified and improved data are then interpreted in terms of two independent models of nuclear reactions, a potential model and an R-matrix approach. This attention to detail was clearly not practical in previous reaction rate evaluations. After combining the results from the two reaction models, we find $^{16}$O(p,$\gamma$)$^{17}$F reaction rate errors of less than 7\% over the entire range of astrophysical interest. In other words, the $^{16}$O(p,$\gamma$)$^{17}$F reaction exhibits now the most precisely determined thermonuclear rates among any charged-particle capture reactions in the $A\geq12$ mass range. Our new results are then incorporated into models of massive AGB stars in order to study the derived oxygen isotopic ratios. Contrary to previous conclusions, we find now that there is presently no clear evidence of a massive AGB star origin for any observed stellar grains.

\begin{acknowledgments} 
The authors would like to thank Claudio Ugalde for helpful discussions, Mark van Raai for providing support with the AGB code, Larry Nittler for providing stellar grain data, and Amanda Karakas for improving the stellar structure input. This work was supported in part by the U.S. Department of Energy under Contract No. DE-FG02-97ER41041 and by the Netherlands Organisation for Scientific Research (NWO). 
\end{acknowledgments}


\begin{thebibliography}{99}
\bibitem{TIL93} D.~R. Tilley, H.~R. Weller and C.~M. Cheves, Nucl. Phys. A 564, 1 (1993).
\bibitem{ROL73} C. Rolfs, Nucl. Phys. A 217, 29 (1973).
\bibitem{ILI07} C. Iliadis, Nuclear Physics of Stars (Wiley-VCH, Weinheim, 2007).
\bibitem{HES58} R.~E. Hester, R.~E. Pixley and W.~A.~S. Lamb, Phys. Rev. 111, 1604 (1958).
\bibitem{TAN59} N. Tanner, Phys. Rev. 114, 1060 (1959).
\bibitem{CHO75} H.~C. Chow, G.~M. Griffiths and T.~H. Hall, Can. J. Phys. 53, 1672 (1975).
\bibitem{MOR97} R. Morlock et al., Phys. Rev. Lett. 79, 3837 (1997).
\bibitem{BEC82} H.~W. Becker et al., Z. Phys. A 305, 319 (1982).
\bibitem{ANG99} C. Angulo et al., Nucl. Phys. A 656, 3 (1999).
\bibitem{Her05} F. Herwig, Ann. Rev. Astr. Astrophys. 43, 435 (2005).
\bibitem{HO04} H.~J. Habing and H. Olofsson, Asymptotic Giant Branch Stars (Springer, Heidelberg, 2004).
\bibitem{LUG05} M. Lugaro, Stardust from Meteorites (World Scientific, Singapore, 2005).
\bibitem{LUG07} M. Lugaro et al., Astron. Astrophys. 461, 657 (2007).
\bibitem{AUD03} G. Audi, A.~H. Wapstra and C. Thibault, Nucl. Phys. A 729, 337 (2003).
\bibitem{MOR96} R. Morlock, Diplom thesis (Universit\"at Stuttgart, 1997).
\bibitem{KOE99} V. K\"olle et al., Nucl. Instr. Meth. A 431, 160 (1999).
\bibitem{KOE98} V. K\"olle, Ph.D. thesis (Universit\"at T\"ubingen, 1997).
\bibitem{WHA} W. Whaling, Kellogg Radiation Laboratory, California Institute of Technology (unpublished).
\bibitem{SRIM} J.F. Ziegler and J.P. Biersack, computer program SRIM, 2003.
\bibitem{CHA07} A. Chafa et al., Phys. Rev. C 75, 035810 (2007).
\bibitem{NEW07} J.~R. Newton et al., Phys. Rev. C 75, 055080 (2007).
\bibitem{BEC07} H.~W. Becker, priv. comm. (2007).
\bibitem{BAD56} M. Bader, R.E. Pixley, F.S. Mozer and W. Whaling, Phys. Rev. 103, 32 (1956).
\bibitem{DAA04} P. Descouvemont,  A. Adahchour,  C. Angulo,  A. Coc, and E. Vangioni-Flam,  At. Data Nucl. Data Tables 88, 203 (2004).
\bibitem{ILI01} C. Iliadis, J.~M. D'Auria, S. Starrfield, W.~J. Thompson and M. Wiescher, Astrophys. J. Suppl. Ser. 134, 151 (2001).
\bibitem{ILI04} C. Iliadis and M. Wiescher, Phys. Rev. C 69, 064305 (2004).
\bibitem{COM01} In fact, the quantity $\mathcal{S}$ denotes the spectroscopic factor, while C is an isospin Clebsch-Gordan coefficient. However, in order to avoid confusion with the astrophysical S-factor, we will refer to $C^2\mathcal{S}$ as the spectroscopic factor.
\bibitem{Kim87}
  K. H. Kim, M. H. Park, and B. T. Kim, \prc  {\bf 35}, 363 (1987).
\bibitem{Mohr93}
  P.~Mohr, H.~Abele, R.~Zwiebel, G.~Staudt, H.~Krauss, H.~Oberhummer,
  A.~Denker, J.~W.~Hammer, G.~Wolf,
  \prc {\bf 48}, 1420 (1993).
\bibitem{DES99} P. Descouvemont and D. Baye, Phys. Rev. C 60, 015803 (1999).
\bibitem{ABE93} H. Abele and G. Staudt, Phys. Rev. C 47, 742 (1993).
\bibitem{LT58} A.M. Lane and  R.G. Thomas, Rev. Mod. Phys. 30, 257 (1958).
\bibitem{AD98} C. Angulo and  P. Descouvemont, Nucl.Phys. A 639, 733 (1998).
\bibitem{BDH98} D. Baye,  P. Descouvemont, and M. Hesse,  Phys. Rev. C 58, 545 (1998).
\bibitem{trigo07} J.M. Trigo-Rodr\'igues, D.A. Garc\'ia-Hern\'andez, M. Lugaro, A.I. Karakas, M. van Raai, P. 
Garc\'ia Lario, and A. Manchado, Meteorit. and Planet. Sci., submitted (2007).
\bibitem{boothroyd95} A. Boothroyd, I.-J. Sackmann, and G.J. Wasserburg, Astrophys. J., 442, 
L21 (1995).
\bibitem{karakas07} A.I. Karakas and J.C. Lattanzio, Publ. Astron. Soc. Australia, in press 
(2007)
\bibitem{vassiliadis93}	E. Vassiliadis and P.R. Wood, Astrophys. J. 413, 641 (1993).
\bibitem{chafa05} A. Chafa et al., Phys. Rev. Lett. 95, 1101 (2005).
\bibitem{fox05} C. Fox et al., Phys. Rev. C 71, 055801 (2005).
\bibitem{ventura05a} P. Ventura and F. D'Antona, Astron. Astrophys. 431, 279 (2005).
\bibitem{ventura05b} P. Ventura and F. D'Antona, Astron. Astrophys. 439, 1075 (2005).
\bibitem{bernatowicz05} T.J. Bernatowicz, O.W. Akande, T.K. Croat, and R. Cowsik, Astrophys. J. 631, 988 (2005).
\bibitem{nuth06} J. A. III Nuth, G.M. Wilkinson, N.M. Johnson, and M. Dwyer, Astrophys. J. 644, 1164 (2006).
\end{thebibliography}
\end{document}